\documentclass[10pt]{article}

\usepackage[english]{babel}
\usepackage{graphicx}
\usepackage{epsfig}
\usepackage{ae}
\usepackage{subfigure}
\usepackage{url}
\usepackage{svn}
\usepackage{amsfonts}
\usepackage{amsmath}
\usepackage{amssymb}
\usepackage{amstext}
\usepackage{amsthm}
\usepackage{xspace}
\usepackage{multirow}
\usepackage{tabularx}
\usepackage{booktabs}
\usepackage{tkz-graph}
\usepackage{multirow}
\usepackage{rotating}

\newcommand{\st}{\mbox{s.t.}}

\renewcommand*{\arraystretch}{1}
\renewcommand{\tabcolsep}{3pt}

\title{Optimal offline virtual network embedding \\ with rent-at-bulk aspects}
\author{S. Coniglio, B. Grimm, A.M.C.A. Koster, M. Tieves, A. Werner\thanks{S. Coniglio, A.M.C.A. Koster, and M. Tieves are with Lehrstuhl~II f\"ur Mathematik, RWTH Aachen University, Germany. B. Grimm and A. Werner are with Konrad-Zuse-Zentrum f\"ur Informationstechnik, Berlin, Germany. This work is supported by the German Federal Ministry of Education and Research, joint project 05M2013 - VINO: Virtual Network Optimization, BMBF grants 05M13PAA and 05M13ZAC.}}
\begin{document}
\maketitle

\begin{abstract}
Network virtualization techniques allow for the coexistence of many virtual networks (VNs) jointly sharing the resources of an underlying substrate network. 
The Virtual Network Embedding problem (VNE) arises when looking for the most profitable set of VNs to embed onto the substrate. 
In this paper, we address the offline version of the problem.
We propose a Mixed-Integer Linear Programming formulation to solve it to optimality which accounts for acceptance and rejection of virtual network requests, allowing for both splittable and unsplittable (single path) routing schemes.
Our formulation also considers a Rent-at-Bulk (RaB) model 
for the rental of substrate capacities
where economies of scale apply.
To better emphasize the importance of RaB, 
we also compare our method to a baseline one which
only takes RaB into account {\em a posteriori}, once a solution to VNE, oblivious to RaB, has been found.
Computational experiments 
show the viability of our approach, stressing the relevance of addressing RaB directly
with an exact formulation.

\end{abstract}


\section{Introduction}
According to much of the recent literature, {\em network virtualization} techniques are becoming one of the distinctive features of the new generation of network architectures~\cite{chowdhury2009network}.
{\em In nuce}, network virtualization consists in decoupling the traditional role of current generation Internet service providers into two independent roles:
management of the physical network and service provisioning.
Two new actors are thus identified: {\em Service Providers} (SPs), who aggregate physical resources so to realize a set of {\em Virtual Networks} (VNs) by which end-user services are provided, and {\em Infrastructure Providers} (InPs), by whom the physical network is managed and from whom physical resources are rented.
%
For a survey on the topic, we refer the reader to~\cite{chowdhury2010survey}.

\begin{figure}[h!]
\begin{center}
 \begin{tikzpicture}[scale=.5]
    \coordinate (s1) at (0, 0) {};
    \coordinate (s2) at (3, 0) {};
    \coordinate (s3) at (5, 0) {};
    \coordinate (s4) at (1.25, 1.25) {};
    \coordinate (s5) at (4.25, 1.25) {};
    \coordinate (s6) at (6.25, 1.25) {};
    \node  at (0, 1) {\textbf{$G^0$}};

    \node (v1) at (2.5, 3) {};
    \node (v2) at (1.5, 4) {};
    \node (v3) at (0.5, 3) {};
    \node  at (3, 3.75) {\textbf{$R^1$}};

    \node (vv1) at (5.5, 4) {};
    \node (vv2) at (6.5, 3.25) {};
    \node (vv3) at (5.5, 2.5) {};
    \node  at (6.75, 4.125) {\textbf{$R^2$}};

    \filldraw (s1) circle (8pt);
    \filldraw (s2) circle (8pt);
    \filldraw (s3) circle (8pt);
    \filldraw (s4) circle (8pt);
    \filldraw (s5) circle (8pt);
    \filldraw (s6) circle (8pt);

    \filldraw (v1) circle (2pt);
    \filldraw (v2) circle (2pt);
    \filldraw (v3) circle (2pt);

    \filldraw (vv1) circle (2pt);
    \filldraw (vv2) circle (2pt);
    \filldraw (vv3) circle (2pt);

    \draw[very thick] (s1) -- (s2)  -- (s3) -- (s6) -- (s5) -- (s4) -- (s1);
    \draw[very thick] (s2) -- (s5);

    \draw[orange,very thick] (v1) -- (v2) -- (v3) -- (v1);
    \draw[cyan,very thick] (vv1) -- (vv2) -- (vv3) -- (vv1);

    \draw[dashed] (v3) -- (s4);
    \draw[dashed] (v2) -- (s4);
    \draw[dashed] (v1) -- (s2);

    \draw[dashed] (vv3) -- (s3);
    \draw[dashed] (vv2) -- (s6);
    \draw[dashed] (vv1) -- (s2);

    \draw[orange,densely dashed] ([yshift=-4pt,xshift=10pt]s4.east) to ([yshift=-4pt,xshift=-10pt]s5.west);
    \draw[orange,densely dashed] ([yshift=-4pt,xshift=-10pt]s5.west) to ([yshift=10pt,xshift=4pt]s2.north);

    \draw[cyan,densely dashed] ([yshift=-4pt,xshift=10pt]s5.east) to ([yshift=-4pt,xshift=-10pt]s6.west);
    \draw[cyan,densely dashed] ([yshift=-4pt,xshift=-10pt]s6.west) to ([yshift=10pt,xshift=4pt]s3.north);
    \draw[cyan,densely dashed] ([yshift=-10pt,xshift=-4pt]s5.south) to ([yshift=4pt,xshift=10pt]s2.north);
    \draw[cyan,densely dashed] ([yshift=4pt,xshift=10pt]s2.east) to ([yshift=4pt,xshift=-10pt]s3.west);

    \filldraw[orange] ([xshift=-5pt]s2.west) arc (180:0:5pt);
    \filldraw[cyan] ([xshift=-5pt]s2.west) arc (180:360:5pt);

    \filldraw[orange] ([xshift=-5pt]s4.west) arc (180:0:5pt);
    \filldraw[orange] ([xshift=-5pt]s4.west) arc (180:360:5pt);

    \filldraw[cyan] ([xshift=-5pt]s3.west) arc (180:360:5pt);
    \filldraw[cyan] ([xshift=-5pt]s6.west) arc (180:360:5pt);
 \end{tikzpicture}
\end{center}
\caption{An embedding of two VNs onto the physical substrate.}\vspace*{-.45cm}
\label{fig}
\end{figure}
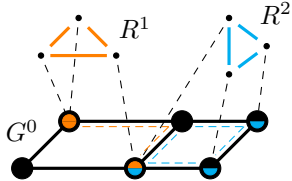
The {\em Virtual Network Embedding} problem (VNE) arises so to efficiently allocate the resources of the physical layer to the various VNs. It calls for a mapping of nodes and links of (a selection of) virtual networks onto nodes and links of the physical layer (see Figure~\ref{fig} for an illustration) so to maximize a profit function.
%
From an optimization point of view, the key aspects of VNE are the following:
\begin{itemize}
\item {\em online} VS {\em offline}: depending on the situation that we consider, VN requests can either arrive dynamically over time or be known {\em a priori};
\item {\em resource requirements}, holding for each VN request (usually of node CPU power and bandwidth);
\item {\em locality requirements}, imposing that a virtual node be mapped only to a specified subset of the physical nodes, (e.g., those belonging to a given region);
\item {\em admission control}, i.e., the possibility of accepting or denying a VN request (e.g., because it is not sufficiently profitable or too resource demanding); 
\item {\em routing schemes}, depending on the architecture and protocols (e.g., splittable routing for UDP traffic or  unsplittable single path routing for MPLS).
\end{itemize}



For both the online and offline cases, with and without admission control, VNE is $\mathcal{NP}$-hard even with splittable routing by reduction from the {\em multiway separator problem}~\cite{andersen2002theoretical}. VNE is still $\mathcal{NP}$-hard even if we ignore the bandwidth requirements due to admitting the {\em generalized assignment problem} as a special case. If admission control has been carried out and the node mapping is given, VNE is polynomially solvable with splittable routing, but it is still $\mathcal{NP}$-hard in the unsplittable case, becoming equivalent to the {\em unsplittable multicommodity flow} problem.

Most of the literature on VNE adopts a two-phase heuristic, carrying out node and link in sequence~\cite{zhu2006algorithms,yu2008rethinking,chowdhury2009virtual}. To our knowledge, the only ``partially exact'' approach is in~\cite{houidi2011virtual}, where a Mixed-Integer Linear Programming (MILP) formulation is proposed to
solve a single step of the online problem
to optimality, i.e., to embed a single VN request or a small group of them. The same formulation is then used, essentially unmodified in its basic structure, for variations of the original problem such as in~\cite{botero2012energy}.

To our knowledge, the offline VNE problem has been almost disregarded in the literature where, in the few instances in which it is mentioned, it is considered only as a mean of comparison for the competitiveness of online algorithms.
In spite of this,
we do believe that the offline problem is of large relevance on its own.
 This is the case, for instance, when the VN requests ask for services (e.g., online gaming networks), 
which, after provisioning starts,  are supposed to last for a very long long time span. Such requests
can be issued long before being actually instantiated, thus giving ample time for offline planning.

Differently from previous work, we also speculate that, in the interaction between SPs and InPs for the rental of physical resources, economies of scale would typically apply. A new aspect of VNE thus arises,
with a direct impact on the profits which are maximized in the problem:
\begin{itemize}
\item {\em rent-at-bulk resources} (RaB): physical resources are rented in bulks, with volume discounts.
\end{itemize}

In this letter, we address the offline version of VNE. We propose a MILP formulation which entails the admission control aspect, allowing for both unsplittable and splittable routing schemes, with a RaB scheme for the rental of physical resources. 
Computational experiments are carried out
on a dataset composed of long-haul and datacenter networks so to assess the viability of our approach. We evaluate both the impact of the two different routing schemes and as well as of the new RaB aspect.
To better illustrate the relevance of the latter, we also compare our solutions to a method which, first, solves a version of VNE which is oblivious to the costs of capacity installation and,
only then, computes ({\em a posteriori}) the corresponding RaB costs.

\section{MILP formulation}

Let $R$ be the set of VN requests.
Let  $G^0=(V^0,A^0)$ be a directed graph representing the physical substrate, with total node and link capacities of  $B_i$ for all $i \in V^0$ and $K_{ij}$ for all $(i,j) \in A^0$. Let $U$ and $Q$ be the set of capacity bulks of different size for nodes and links.
Let $\alpha^u$ and $\beta^q$, for $u\in U$ and $q \in Q$, be the RaB costs for nodes and links. Economies of scales dictate $\frac{\alpha^{u_i}}{u_i} \geq \frac{\alpha^{u_j}}{u_j}$ and $\frac{\beta^{q_i}}{q_i} \geq \frac{\beta^{q_i}}{q_j}$ for all $u_i, u_j \in U$ with $u_i \leq u_j$ and $q_i, q_j \in Q$ with $q_i \leq q_j$, i.e., decreasing unit costs for larger bulks of rented capacity.
For each request $r \in R$, let $V^r$ be the set of virtual nodes, with node requirements $t^r_i$ for each $i \in V^r$. Virtual arcs are implicitly represented via the
traffic matrix $D^r \in \mathbb{R}^{|V^r| \times |V^r|}_+$, where each component $d^r_{vw}$ is a demand between two virtual nodes $v,w \in V^r$. For each $r \in R$ and $v \in V^r$, let $V^0(r,v)$ denote the set of physical nodes on which the virtual node~$v$ can be mapped (due to locality restrictions).
%
For each $r \in R$, let $p^r \geq 0$ be the corresponding profit. 

Let the variable $y^r \in \{0,1\}$ take value $1$ if the request of index $r \in R$ is accepted and $0$ otherwise. Let $x_{vi}^{r} \in \{0,1\}$ be equal to $1$ if the virtual node $v \in V^r$ of request $r \in R$ is mapped onto the physical node $i \in V^0(r,v)$ and to $0$ otherwise. Assuming, for now, an unsplittable routing, let~$f_{ij}^{r,vw}$ take value $1$ if traffic between the two virtual nodes $v,w \in V^r$, for request $r \in R$, is routed over the arc $(i,j) \in A^0$ and $0$ otherwise. This way, we have a multicommodity flow with a commodity per triple $r,v,w$.
Let the integer variables $g_i^u, h_{ij}^q$ denote the amount of bulk of capacity
$u \in U$ and $q \in Q$ rented on the physical node~$i$ and physical link~$(i,j)$, respectively. Our MILP formulation reads:

\small \vspace*{-.5cm}
\begin{align}
\max \;    & \sum_{r \in R} p^r y^r - \sum_{u \in U} \alpha^u \sum_{i \in V^0} g_i^u - \sum_{q \in Q} \beta^q \sum_{(i,j) \in A^0} h_{ij}^q\\
\label{mip:nodemapping}  \st & \sum_{i \in V^0(r,v)}  x_{vi}^r = y^r
  \hspace{2.2cm} \forall r \in R, v \in V^r\\
\label{mip:nodecapacity}     & \sum_{r\in R}\sum_{v \in V^r} t^r_v x_{vi}^r \leq \sum_{u \in U} u \, g_i^u
  \hspace{1.8cm} \forall i \in V^0\\
\label{mip:linkcapacity}     & \sum_{r \in R} \sum_{v, w \in V^r} d^r_{vw} f_{ij}^{r,vw} \leq \sum_{q \in Q} q \, h_{ij}^q
  \hspace{0.4cm} \forall (i,j) \in A^0\\
\label{mip:nodebulk}     & \sum_{u\in U} u \, g_i^u \leq B_i
  \hspace{3.8cm} \forall i \in V^0\\
\label{mip:linkbulk}     & \sum_{q\in Q} q \, h_{ij}^q \leq K_{ij}
  \hspace{3.0cm} \forall (i,j) \in A^0\\
\nonumber                    & \sum_{(i,j) \in \delta^+(i)} f_{ij}^{r,vw} - \sum_{(j,i) \in \delta^-(i)} f_{ji}^{r,vw} = x_{vi}^r - x_{wi}^r\\
\label{mip:flow}             &
  \hspace{3.3cm} \forall r \in R, v,w \in V^r, i \in V^0\\
\label{mip:vars}             & y^r,x_{vi}^r, f_{ij}^{r,vw} \in \{0,1\}, g_i^u,h_{ij}^q \in \mathbb{Z}_+.
\end{align}
\normalsize
Constraints~\eqref{mip:nodemapping} enforce that each virtual node be mapped onto a single physical node
meeting its locality requirements
if the corresponding request is accepted, and to none otherwise. Constraints~\eqref{mip:nodecapacity} and~\eqref{mip:linkcapacity} guarantee that the physical node and link capacity that is used does not exceed the rented one, while Constraints~\eqref{mip:nodebulk} and~\eqref{mip:linkbulk} impose that no more than the total available physical capacity be rented.
Constraints~\eqref{mip:flow} are multicommodity flow balance constraints which, differently from those of a standard multicommodity flow problem,  have a variable right-hand side. This way, physical node $i$ acts as source node if $x_{vi}^r = 1$ and $x_{wi}^r = 0$, as sink node if $x_{vi}^r = 0$ and $x_{wi}^r = 1$, and as a regular intermediate node if $x_{vi}^r = x_{wi}^r = 0$. If $x_{vi}^r = x_{wi}^r = 1$, then the two virtual nodes $v,w$ are mapped onto the same physical node
(the so-called {\em co-location})
and, hence, their traffic demand $d_{vw}^r$ does not need to be routed. 
Constraints~\eqref{mip:vars} denote the nature of the variables. 
Due to variable $f_{ij}^{r,vw}$ being defined
as binary, a single path
routing is enforced. A splittable routing can be obtained by 
relaxing $f_{ij}^{r,vw}$ to be 
 in $[0,1]$ which, this way,  
denotes the fraction of flow of request $r \in R$ between $v,w \in V^r$ that is allocated on the arc $(i,j)$.

Note that, although the formulation proposed in~\cite{houidi2011virtual} can be used in the offline setting as well, it does not allow for admission control, it lacks the RaB scheme for physical resources, and it only allows for a splittable routing. Since such formulation aggregates, for each VN, all the flow between pairs of virtual nodes that are mapped on the same pair of physical nodes, by imposing integrality on the corresponding flow variables we would introduce extra routing constraints, thus incorrectly forcing all such heterogeneous flows
to share the same physical path.

\section{Computational comparison}\label{sec:results}

In this section, we report on a set of computational experiments carried out to assess the effectiveness of our MILP formulation when solved via a commercial branch-and-bound code, as well as to evaluate the impact of the different aspects of the problem on its solvability and on the quality of its solutions.
We adopt the state-of-the-art MILP solver CPLEX 12.6 with default parameters,
halting the execution of CPLEX as soon as a solution with an optimality gap $\leq 1\%$ is found.
%
The experiments are run on a default desktop computer
with an Intel~i7 processor and 16 GB of RAM, within a time limit of $3600$ seconds.

%

We consider two types of topologies for the physical networks. We take five SNDlib~\cite{orlowski2010sndlib} instances ({\tt abilene}, {\tt atlanta}, {\tt france}, {\tt germany50}, {\tt nobel-eu}) to model ``long-haul networks'' representing 
large-scale backbone networks with geographically scattered data centers, each represented as a single aggregate node,
and five {\em transition-stub} instances with $|V^0|$ and $|A^0|$ in $(13, 14, 23, 31, 45)$, $(30,48,60,96,148)$, generated along the procedure in~\cite{zegura1996model}, representing ``data-center networks'', that is, clusters of computers connected via short-haul links. 
For each undirected link in the original instance, two antiparallel arcs are introduced (with possibly different capacities), thus creating directed networks. Virtual node and link capacities are randomly generated with values equal to 5, 10, 50, 500, chosen with probability 0.1, 0.4, 0.4, 0.1, respectively.
From each topology, we generate two different physical network instances by varying the random seed, thus creating a total of 20 physical network instances. We consider bulks of size 1, 10, 100, with costs 1, 5, 25, for both nodes and links.

We couple each of these 20 
substrates with a set of
10, 15, 20, 25 virtual network requests~({\em Req}). Each VN has a random number of nodes between 2 and 10, with a traffic demand $D^r$ with a density of 50\%. Node and link demand requirements are randomly sampled as in the physical layer, with a scaling of
0.3, 0.4, 0.5 ({\em Scal}).
Embedding profits are set to 500. The resulting data set is composed of
240
instances, 10 per configuration of the {\em Req} and {\em Scal} parameters (thus representing problems with different {\em load levels}) and type of substrate.
%
As to the locality aspect, each set $V^0(r,v)$ is constructed by first sampling uniformly at random a cardinality factor $\gamma^r_v$ from the interval $\left[\frac{1}{2}, 1 \right]$. Afterwards, every node $v \in V^0$ is added to $V^0(r,v)$ with probability $\gamma^r_v$, uniformly at random.

The computational results for VNE with RaB are summarized in the first half of Table~\ref{tab_all},
for both splittable and unsplittable routing schemes.
To show the impact of RaB on the solvability of the problem, in the second half of the table we relax the integrality constraints on the variables $g_i^u$ and~$h_{ij}^q$, thus simulating the case of a
standard (constant) price for embedding, proportional to resource consumption. Note that, this way, the cost of capacity installation is underestimated as any optimal solution will only include fractional amounts of the cheapest bulk. Further, to stress that, in an RaB scenario, RaB cost are relevant and should be accounted for directly in the model, we introduce a baseline method (a heuristic for the VNE problem with RaB) which, based on the solution without RaB, computes only {\em a-posteriori} the corresponding RaB costs for capacity installation.



Each row reports data averaged (in arithmetic mean) over the 10 instances
with the same parameter values {\em Req} and {\em Scal}.
%
The total profit for the embedded VNs in the best feasible solution that is found (averaged over the 10 instances per row) is reported in the {\em Profit} column.
The columns \# and {\em Time} show the number of instances that are solved to optimality within the time limit and the corresponding average computing time. The {\em Gap} column reports the average integrality gap for the  instances that are not solved within the time limit (or 0 if all of them are solved). The percentage difference between the exact and the baseline methods is reported in the {\em Impr} column.
%
Average data are reported in boldface.


The comparison between our exact approach and the baseline method indicates that the former can lead to much more profitable solutions than the latter, with an average improvement of 23.4\% which, after a closer inspection, can be found as high as 50\% on 12.5\% of the cases. Note that the data center instances are harder to solve than the long-haul network ones, yielding, on average, larger computing times, larger gaps, and fewer optimal solutions. In spite of this, we still obtain reasonably small gaps (11.5\% on average) and, most interestingly, an even larger improvement w.r.t. the baseline, equal to 27.3\% on average for the two routing schemes.

We now focus on the differences between the two routing schemes over both types of instances.
We remark that, with splittable routing, we can
achieve larger profits since the corresponding formulation is a partial relaxation of the formulation for the unsplittable case (obtained by relaxing the integrality of the flow variables). In spite of this observation, the table shows that
the profits between the two schemes differ, on average, by no more than 2\% (both with or without RaB), thus indicating that, in practice, the option of a splittable flow does not provide any advantage.
What is more, for splittable routing, we register an increase (w.r.t. the unsplittable case) in the average computing time (for the instances that are solved to optimality within the time limit) of a factor of 1.07 with RaB and of 1.2 without it, as well as an increase in the number of unsolved instances by 16 with RaB, going up to 23 without it.
This
outcome is, most likely,
a consequence of the splittable case yielding a less structured problem, for which finding good quality feasible solutions via the primal heuristics called by the MILP solver is not as easy as for the single path case, where all the variables are binary, as well as due to the more effective presolve phase that the solver can carry out for pure binary problems.


When removing RaB, we observe larger gaps for the unsolved instances, with an increase, for splittable and unsplittable routing, of 3.6 and 1.3 percentage points. Interestingly though, we observe, on average, a substantial decrease in computing time of 53\% and 59\%, for the two schemes, as well as an increase in the average number of instances solved to optimality by, respectively, 2.9 and 3.2.
This is not surprising since RaB introduces a network design aspect, making VNE more similar to classical network design problems which are typically hard to solve. Nevertheless, as shown via the comparison with the baseline method, such aspect should not be neglected.

\begin{figure}
\small
\begin{center}
{\scriptsize
\begin{tabular}{cc}
\includegraphics[scale=0.45]{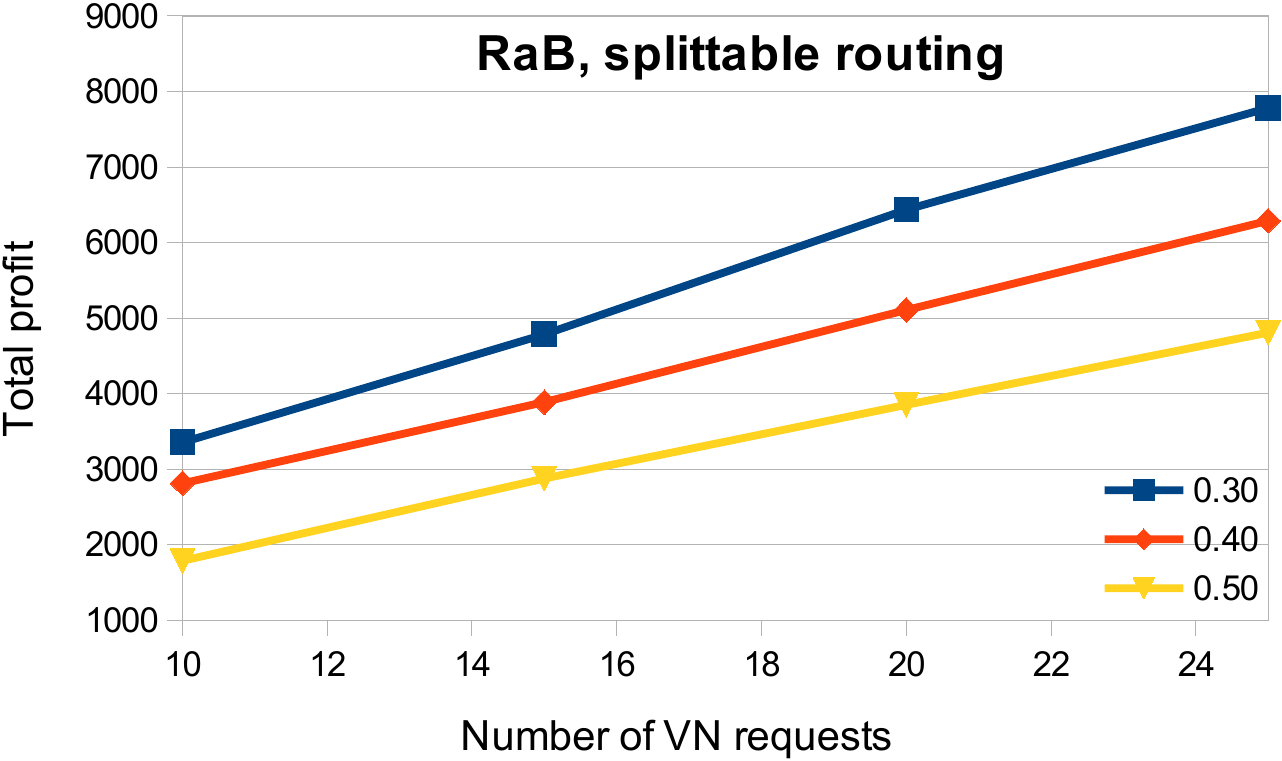} & \includegraphics[scale=0.45]{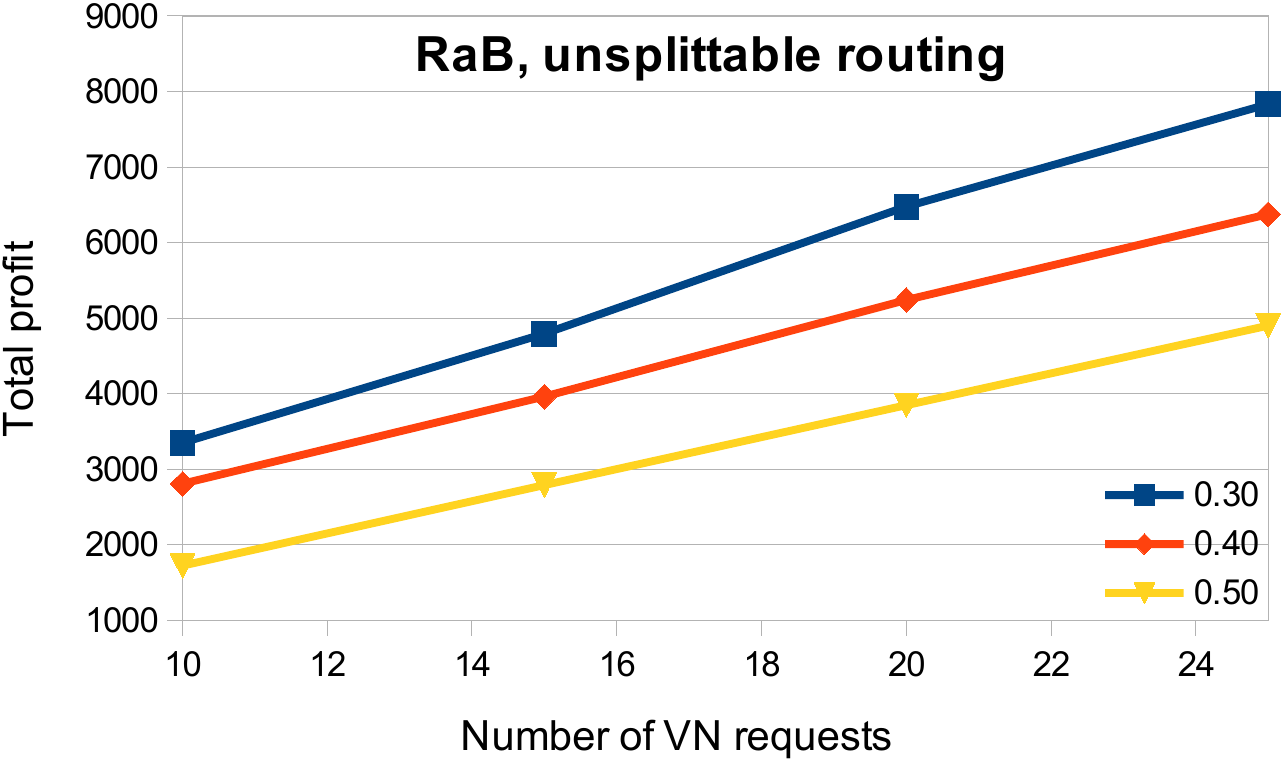}\\
\includegraphics[scale=0.45]{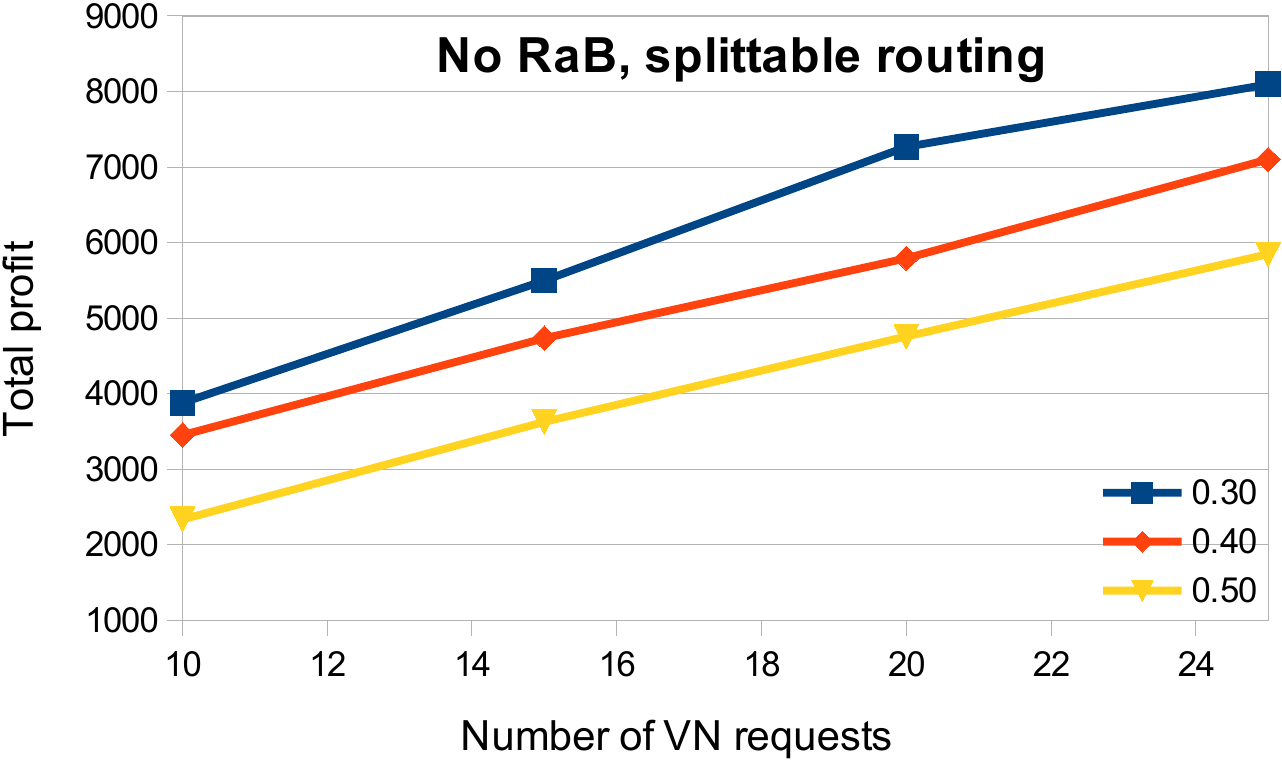} & \includegraphics[scale=0.45]{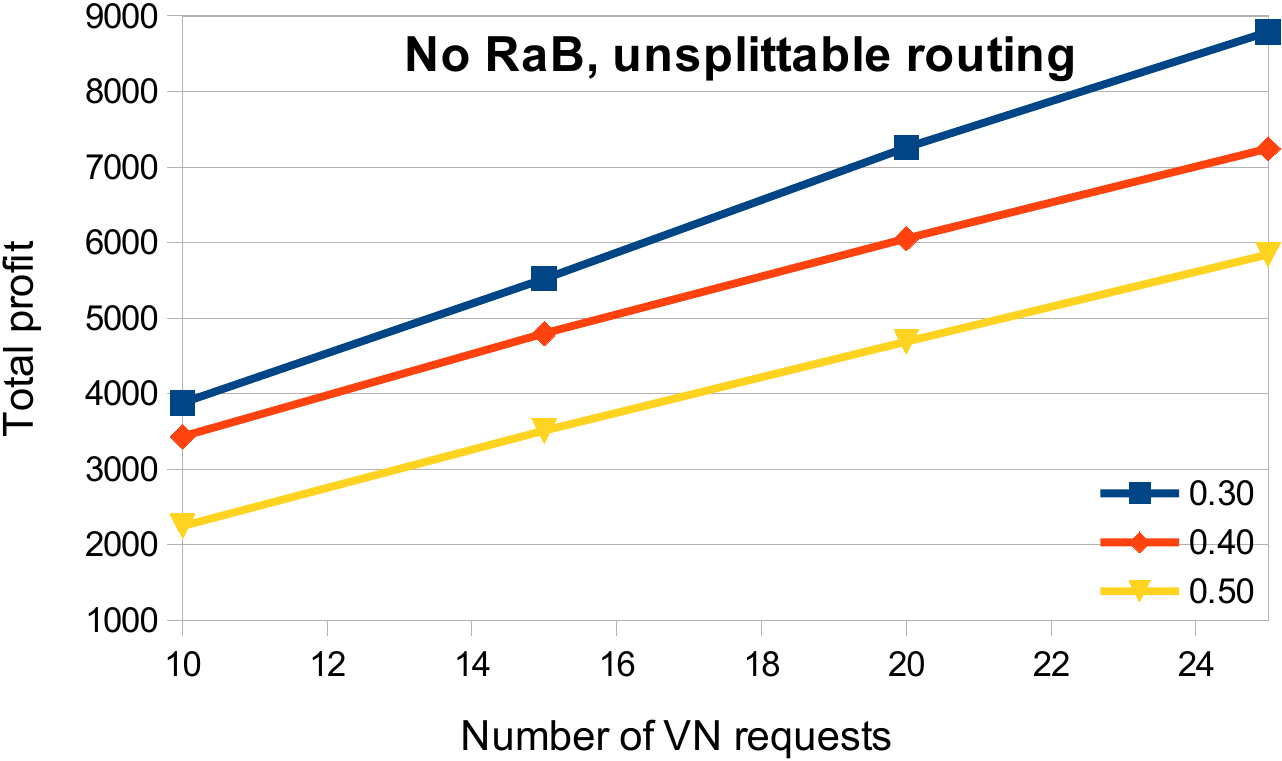}\\
\end{tabular}
}
\end{center}
\caption{Average profits plotted against the number of requests, with a scaling of 0.3, 0.4, 0.5, with splittable and unsplittable routing schemes and with and without RaB.}
\label{fig-charts}
\end{figure}

To conclude, in Figure~\ref{fig-charts} we compare the profits for the embedded VNs requests in the four configurations (splittable or unsplittable routing, with or without RaB), for different values of the scaling parameter {\em Scal} (thus progressively increasing the {\em load level} of the instances).
Interestingly, the charts report the same qualitative growth for all the four cases. 
Nevertheless, we observe that the profits for the case without RaB are larger than those obtained for RaB, a feature which better underlines how, for a practical scenario with RaB, the model without it underestimates, often substantially, the actual costs.

\begin{sidewaystable}[ht!]
\centering
\caption{Results for splittable and single path
routing schemes, with and without rent-at-bulk.}
\renewcommand*{\arraystretch}{0.9}
\renewcommand{\tabcolsep}{1.5pt}
\begin{tabular}{rr|rrrr|rr|rrrr|rr||rrrr|rrrr}

 &  & \multicolumn{ 12}{c||}{\textbf{RaB}} & \multicolumn{ 8}{c}{\textbf{No RaB}} \\ 
\hline
 &  & \multicolumn{ 6}{c|}{\textbf{Splittable routing}} & \multicolumn{ 6}{c||}{\textbf{Unsplittable routing}} & \multicolumn{ 4}{c|}{\textbf{Splittable routing}} & \multicolumn{ 4}{c}{\textbf{Unsplittable routing}} \\ 
 &  & \multicolumn{ 4}{c}{\textbf{Exact}} & \multicolumn{ 2}{c|}{\textbf{Baseline}} & \multicolumn{ 4}{c}{\textbf{Exact}} & \multicolumn{ 2}{c||}{\textbf{Baseline}} & \multicolumn{ 4}{c|}{\textbf{Exact}} & \multicolumn{ 4}{c}{\textbf{Exact}} \\ 
\hline
\textbf{Req} & \textbf{Scal} & \textbf{Profit} & \textbf{\#} & \textbf{Time} & \textbf{Gap} & \textbf{Profit} & \textbf{Impr} & \textbf{Profit} & \textbf{\#} & \textbf{Time} & \textbf{Gap} & \textbf{Profit} & \textbf{Impr} & \textbf{Profit} & \textbf{\#} & \textbf{Time} & \textbf{Gap} & \textbf{Profit} & \textbf{\#} & \textbf{Time} & \textbf{Gap} \\
\hline
\multirow{12}{*}{\rotatebox{90}{{\bf Long-haul networks}}}
10 & 0.3 & 3367 & 9 & 233.5 & 9.4 & 2939 & 14.5 & 3366 & 9 & 96.7 & 9.4 & 3065 & 9.8 & 3670 & 9 & 2.4 & 11.1 & 3672 & 9 & 0.8 & 11.1 \\ 
10 & 0.4 & 2926 & 10 & 382.3 & 0.0 & 2335 & 25.3 & 2913 & 10 & 228.4 & 0.0 & 2343 & 24.3 & 3322 & 10 & 9.1 & 0.0 & 3267 & 10 & 2.2 & 0.0 \\ 
10 & 0.5 & 1934 & 10 & 126.6 & 0.0 & 1227 & 57.5 & 1903 & 10 & 187.7 & 0.0 & 1363 & 39.6 & 2305 & 10 & 46.4 & 0.0 & 2297 & 10 & 2.1 & 0.0 \\ 
15 & 0.3 & 4978 & 8 & 568.6 & 4.1 & 4205 & 18.4 & 4951 & 9 & 340.4 & 3.1 & 4353 & 13.7 & 5364 & 10 & 30.3 & 0.0 & 5314 & 9 & 2.8 & 7.7 \\ 
15 & 0.4 & 4116 & 9 & 801.4 & 8.7 & 3353 & 22.8 & 4113 & 9 & 172.3 & 2.1 & 3444 & 19.4 & 4575 & 9 & 42.2 & 15.4 & 4621 & 10 & 73.7 & 0.0 \\ 
15 & 0.5 & 3149 & 7 & 441.3 & 3.0 & 2576 & 22.2 & 3133 & 10 & 625.3 & 0.0 & 2542 & 23.3 & 3620 & 9 & 247.3 & 8.3 & 3621 & 10 & 4.1 & 0.0 \\ 
20 & 0.3 & 6661 & 8 & 650.6 & 1.1 & 5977 & 11.4 & 6658 & 10 & 442.5 & 0.0 & 5958 & 11.7 & 7024 & 10 & 85.0 & 0.0 & 7021 & 10 & 26.1 & 0.0 \\ 
20 & 0.4 & 5361 & 6 & 493.8 & 15.6 & 4714 & 13.7 & 5492 & 7 & 510.4 & 11.1 & 4656 & 18.0 & 5669 & 7 & 40.5 & 42.4 & 5983 & 8 & 33.7 & 18.0 \\ 
20 & 0.5 & 4045 & 6 & 618.1 & 40.5 & 3489 & 15.9 & 4221 & 7 & 431.2 & 13.7 & 3338 & 26.5 & 4785 & 7 & 290.8 & 49.7 & 4866 & 8 & 239.4 & 16.7 \\ 
25 & 0.3 & 8353 & 5 & 855.3 & 2.7 & 7659 & 9.1 & 8341 & 6 & 507.5 & 2.2 & 7480 & 11.5 & 7804 & 9 & 84.7 & 10.7 & 8678 & 10 & 42.7 & 0.0 \\ 
25 & 0.4 & 6904 & 6 & 759.4 & 9.7 & 6058 & 14.0 & 7008 & 7 & 383.1 & 6.3 & 6284 & 11.5 & 7596 & 8 & 292.2 & 24.7 & 7659 & 10 & 335.8 & 0.0 \\ 
25 & 0.5 & 5135 & 3 & 1148.7 & 11.9 & 4494 & 14.3 & 5298 & 4 & 452.5 & 5.2 & 4428 & 19.6 & 6192 & 7 & 477.2 & 19.4 & 6041 & 9 & 349.9 & 33.3 \\ 
\textbf{Avg} &  & \textbf{4744} & \textbf{7.3} & \textbf{590.0} & \textbf{8.9} & \textbf{4085} & \textbf{19.9} & \textbf{4783} & \textbf{8.2} & \textbf{364.8} & \textbf{4.4} & \textbf{4105} & \textbf{19.1} & \textbf{5160} & \textbf{8.8} & \textbf{137.3} & \textbf{15.1} & \textbf{5253.0} & \textbf{9.4} & \textbf{92.8} & \textbf{7.2} \\ 
\hline
\multirow{12}{*}{\rotatebox{90}{{\bf Data-center networks}}}
10 & 0.3 & 3349 & 2 & 788.3 & 5.3 & 2549 & 31.4 & 3332 & 2 & 1462.5 & 5.9 & 2647 & 25.9 & 4086 & 9 & 90.6 & 11.1 & 4084 & 9 & 91.0 & 11.1 \\ 
10 & 0.4 & 2702 & 2 & 95.1 & 6.3 & 1822 & 48.3 & 2701 & 2 & 126.2 & 6.2 & 1898 & 42.3 & 3583 & 8 & 782.5 & 11.8 & 3595 & 8 & 939.5 & 11.8 \\ 
10 & 0.5 & 1636 & 3 & 27.4 & 4.3 & 918 & 78.3 & 1540 & 6 & 493.2 & 5.1 & 900 & 71.0 & 2368 & 10 & 113.1 & 0.0 & 2202 & 10 & 3.3 & 0.0 \\ 
15 & 0.3 & 4598 & 1 & 802.7 & 11.7 & 3784 & 21.5 & 4636 & 2 & 2218.5 & 11.7 & 3862 & 20.0 & 5635 & 7 & 255.6 & 13.0 & 5726 & 8 & 276.9 & 7.4 \\ 
15 & 0.4 & 3660 & 2 & 861.2 & 20.5 & 2975 & 23.0 & 3807 & 2 & 703.8 & 14.5 & 2932 & 29.8 & 4897 & 5 & 511.9 & 16.6 & 4973 & 6 & 512.7 & 15.6 \\ 
15 & 0.5 & 2605 & 3 & 293.2 & 9.5 & 1889 & 37.9 & 2450 & 4 & 248.3 & 6.3 & 1776 & 38.0 & 3633 & 7 & 426.0 & 13.5 & 3399 & 10 & 230.8 & 0.0 \\ 
20 & 0.3 & 6216 & 1 & 351.8 & 12.1 & 5600 & 11.0 & 6294 & 1 & 433.9 & 10.4 & 5525 & 13.9 & 7513 & 7 & 338.5 & 5.6 & 7499 & 8 & 117.1 & 8.5 \\ 
20 & 0.4 & 4855 & 2 & 794.3 & 19.5 & 4354 & 11.5 & 4991 & 2 & 876.9 & 15.5 & 4365 & 14.3 & 5914 & 4 & 928.9 & 22.0 & 6128 & 4 & 471.0 & 15.3 \\ 
20 & 0.5 & 3663 & 3 & 480.1 & 13.3 & 2894 & 26.6 & 3473 & 3 & 227.2 & 8.7 & 2677 & 29.8 & 4730 & 6 & 185.1 & 18.8 & 4512 & 9 & 157.4 & 13.9 \\ 
25 & 0.3 & 7217 & 1 & 1553.1 & 15.1 & 6550 & 10.2 & 7329 & 1 & 1145.3 & 12.9 & 6778 & 8.1 & 8392 & 6 & 668.0 & 19.1 & 8904 & 8 & 1144.4 & 7.4 \\ 
25 & 0.4 & 5677 & 1 & 246.3 & 20.3 & 5168 & 9.8 & 5740 & 1 & 243.7 & 18.9 & 5202 & 10.3 & 6608 & 2 & 153.5 & 23.8 & 6831 & 2 & 43.8 & 19.0 \\ 
25 & 0.5 & 4482 & 3 & 382.9 & 15.9 & 3677 & 21.9 & 4502 & 3 & 316.4 & 8.4 & 3721 & 21.0 & 5502 & 4 & 295.2 & 16.7 & 5647 & 8 & 252.6 & 11.4 \\ 
\textbf{Avg} &  & \textbf{4222} & \textbf{2.0} & \textbf{556.4} & \textbf{12.8} & \textbf{3515} & \textbf{27.6} & \textbf{4233} & \textbf{2.4} & \textbf{708.0} & \textbf{10.4} & \textbf{3524} & \textbf{27.0} & \textbf{5238} & \textbf{6.3} & \textbf{395.8} & \textbf{14.3} & \textbf{5292} & \textbf{7.5} & \textbf{353.4} & \textbf{10.1} \\ 
\hline
\multicolumn{2}{l|}{{\bf Tot avg}} & \textbf{4483} & \textbf{4.6} & \textbf{573.2} & \textbf{10.9} & \textbf{3800} & \textbf{23.8} & \textbf{4508} & \textbf{5.3} & \textbf{536.4} & \textbf{7.4} & \textbf{3814} & \textbf{23.1} & \textbf{5199} & \textbf{7.5} & \textbf{266.5} & \textbf{14.7} & \textbf{5272} & \textbf{8.5} & \textbf{223.1} & \textbf{8.7} \\ 
\end{tabular}

\label{tab_all}
\vspace*{-0.5cm}
\end{sidewaystable}

\section{Conclusions and further research}\label{sec:conclusions}

We have presented
an exact MILP
formulation for the offline VNE problem, calling for the selection of the most profitable subset of VNs to embed onto the substrate.
Differently from previous work, our formulation is suitable for both splittable and single path routing schemes
and introduces a new aspect of VNE, the RaB scheme for the rental of capacities from the
substrate. Computational experiments have shown the importance of RaB when comparing to a method which neglect it, as well as the overall viability of our exact MILP approach.
Future work includes the study of the polyhedral structure of the problem, so as to introduce tighter constraints yielding better bounds and, overall, a faster solution process.

\bibliography{references}
\bibliographystyle{unsrt}
\addcontentsline{toc}{chapter}{biblio}



\end{document}